\documentclass[journal,twoside,web]{ieeecolor}
\usepackage{generic}
\usepackage{cite}
\usepackage{amsmath,amssymb,amsfonts}

\usepackage{graphicx}
\usepackage{algorithm}
\usepackage{algpseudocode}

\usepackage{booktabs}
\usepackage{threeparttable}
\usepackage{xurl}
\usepackage{hyperref}

\hypersetup{hidelinks=true}
\usepackage{textcomp}
\def\BibTeX{{\rm B\kern-.05em{\sc i\kern-.025em b}\kern-.08em
    T\kern-.1667em\lower.7ex\hbox{E}\kern-.125emX}}
\providecommand{\refname}{References}
\begin{document}
\title{Sparse-View 3D Coronary Reconstruction from Vessel Segmentations Using a Diffusion Prior}
\author{Tianyuan~Wang, Kenan~Niu, and
Stefano~Stramigioli,~\mbox{\IEEEmembership{Fellow,~IEEE}}
\thanks{This research was financed by the EFRO OOST fund.}
\thanks{Tianyuan Wang, Kenan Niu, and Stefano Stramigioli are with
the Robotics and Mechatronics group, University of Twente,
7500 AE Enschede, The Netherlands
(e-mail: tianyuan.wang@utwente.nl, k.niu@utwente.nl, s.stramigioli@utwente.nl).}
}

\maketitle

\begin{abstract}
Reconstructing three-dimensional (3D) coronary anatomy from sparse angiographic views could support percutaneous coronary intervention, but limited angular coverage and imperfect vessel segmentations introduce substantial geometric ambiguity. We propose a vessel-specific reconstruction framework combining a learned 3D diffusion prior, visual-hull initialization, and projection-based data consistency. The unconditional diffusion model is trained on coronary vessel masks without paired 2D–3D data and reused across different view counts without retraining. During inference, a scaled visual hull initializes reconstruction, while projection-based updates constrain diffusion refinement using patient-specific vessel observations. We evaluated the framework on simulated angiograms derived from ImageCAS computed tomography angiography volumes using two, four, and eight views at two contrast levels. Across 50 evaluation coronary-tree cases, the proposed method achieved higher mean Dice and clDice than visual hull, deep image prior (DIP), and hash-grid implicit neural representation (INR) at all view counts under reduced contrast. With eight views, improvements over DIP, the best-performing baseline at lower contrast, were approximately 0.13 in Dice and 0.22 in clDice. At higher contrast, performance remained competitive with hash-grid INR, the best-performing eight-view baseline. The preferred diffusion starting timestep depended on observation quality and view count, with excessive refinement reducing accuracy at higher contrast. Projection guidance improved both metrics across all conditions. Averaging stochastic reconstructions further improved accuracy, while sample disagreement localized errors with average-precision gains of 0.14–0.21 over error prevalence. These findings support diffusion-prior reconstruction from sparse vessel observations, although clinical validation remains necessary.
\end{abstract}


\section{Introduction}
\label{sec:introduction}
\IEEEPARstart{P}{ercutaneous} coronary intervention (PCI) is primarily performed under two-dimensional (2D) X-ray angiography and fluoroscopy, where the interventionalist navigates guidewires and catheters through a complex three-dimensional (3D) coronary anatomy based on projected images \cite{pore2023autonomous}. Although the C-arm can be rotated to acquire the coronary tree from different viewing angles, each projection inevitably loses depth information and may suffer from vessel overlap and foreshortening. In clinical practice, these ambiguities are largely resolved through the operator's experience and by acquiring additional projections when necessary. Similar ambiguities pose an even greater challenge for image-guided robotic PCI, where a robotic system must infer the spatial relationship between the device and vascular anatomy from limited 2D observations. Since different viewing angles are already accessible with a rotational C-arm, an attractive alternative is to exploit a small number of angiographic views to recover a 3D representation of the coronary vasculature, thereby providing more explicit spatial information for either human or robotic navigation.

Unlike conventional computed tomography (CT), PCI guidance does not necessarily require reconstruction of the full attenuation distribution within the imaged volume. For navigation, the primary structures of interest are the coronary vessels themselves. This motivates a task-specific reconstruction formulation in which vessel-related information is first extracted from a small number of X-ray projections through segmentation and subsequently used to infer the corresponding 3D coronary geometry \cite{ravigopal2022fluoroscopic}. Such a formulation focuses reconstruction on vascular structures, but recovering a complex vascular tree from only a few segmented projections remains severely ill-posed. In particular, multiple 3D vessel configurations may produce similar 2D segmented observations, and distal branches or overlapping vessels may not be uniquely resolved from the available views. Anatomical prior knowledge can therefore help constrain the reconstruction and reduce these ambiguities.

Generative models provide a promising mechanism for incorporating such prior knowledge \cite{chung2022diffusion}. In particular, diffusion models can learn the distribution of anatomically plausible 3D coronary structures from a population of vessel geometries and can therefore constrain sparse-view reconstruction toward realistic vascular anatomies. From an inverse-problem perspective, the reconstruction can be regarded as balancing two sources of information \cite{chung2023solving}: patient-specific observations derived from the acquired X-ray images and a learned anatomical prior provided by the diffusion model. A strong prior can compensate for information that is absent from sparse projections, but it must remain sufficiently constrained by the observations to preserve the patient-specific vessel geometry. The appropriate balance between these two components therefore becomes a central issue in diffusion-prior reconstruction.

This balance becomes particularly important when the observations themselves are uncertain. In a vessel-specific reconstruction pipeline, the X-ray projections are first converted into vessel observations by a segmentation model. However, segmentation quality depends strongly on the visibility of the vessels in the original angiographic images. Reduced vessel contrast, noise, overlap, or weak visualization of small branches may lead to missing vessels, inaccurate boundaries, or false-positive structures. If these imperfect segmentations are imposed as strong deterministic constraints, their errors can directly propagate into the reconstructed 3D vessel tree and prevent the learned prior from correcting them. Conversely, if the observations are weighted too weakly, the reconstruction may follow the learned population prior while losing important patient-specific information.

In this work, we investigate sparse-view coronary reconstruction
by combining a pretrained 3D anatomical diffusion prior with
soft visual-hull initialization and projection-based data
consistency. Rather than treating segmented vessels as exact
geometric constraints throughout reconstruction, the visual hull
provides an initial spatial estimate, while projection-based
updates encourage agreement with the observed vessel structures
during diffusion refinement. We systematically vary vessel
contrast and the number of projection views to investigate how
observation quality and angular sparsity affect reconstruction
performance and the preferred diffusion starting timestep.
This allows us to examine the tradeoff between preserving
measurement-derived geometry and allowing refinement by the
learned anatomical prior. The proposed framework aims to
recover patient-specific 3D coronary geometry from a small
number of angiographic views, with potential applications in
human or robotic PCI guidance.

The main contributions of this work are as follows:

1. We formulate sparse-view coronary reconstruction as a vessel-specific inverse problem and introduce a framework that combines a learned anatomical diffusion prior with soft visual-hull initialization and projection-based data consistency.

2. We systematically investigate how vessel contrast and the number of projection views affect the reconstruction performance of the proposed method and baseline approaches when using imperfect vessel segmentations.

3. We demonstrate that averaging reconstruction samples generated using different noise realizations improves accuracy relative to single-sample reconstruction, while disagreement among samples provides a spatial indicator of potentially unreliable regions.

\section{Related Work}
\label{sec:related_work}

\subsection{Sparse-View Coronary Artery Reconstruction}
\label{sec:related_coronary_reconstruction}
Tomographic coronary reconstruction generally requires more projection
views and broader angular coverage than geometry-based approaches
\cite{ccimen2016reconstruction}. Banerjee et al. combined rigid and
non-rigid motion correction with iterative point-cloud reconstruction
to recover coronary vessel geometry from as few as two non-simultaneous
angiographic views \cite{banerjee2019point}. They evaluated centerline
accuracy using two-dimensional reprojection errors and compared lumen
cross-sectional areas with optical coherence tomography measurements.
However, the geometric accuracy of the complete reconstructed coronary
tree was not directly quantified against an independent
three-dimensional reference.

Supervised learning methods can replace part of this explicit geometric
reasoning with anatomical information learned from training data.
Iyer \emph{et al.} trained a multi-stage neural network on synthetic
coronary trees and their corresponding projections to estimate vessel
centerlines and radii without requiring acquisition parameters as
network inputs \cite{iyer2023multi}. 3DAngioNet constructs a coarse
vessel mesh from learned segmentations of two angiographic views,
calibrated acquisition geometry, and segment endpoints identified by
a clinician. The mesh is subsequently refined by a graph convolutional
network using image features sampled at the projected mesh vertices
\cite{bransby20233d}. DeepCA is trained on projections simulated from
3D vessel masks, with motion introduced between views, to reconstruct
coronary trees from two non-simultaneous angiograms with implicit
compensation for cardiac and respiratory motion \cite{wang2025deepca}.
However, these supervised approaches learn reconstruction mappings
from paired 2D--3D data under the acquisition conditions represented
during training. Their generalization can therefore depend on the
quality and diversity of the training data, the range of input-view
configurations considered, and the similarity between training and
clinical data distributions.

Patient-specific reconstruction methods instead optimize a
three-dimensional representation directly against the available
projections, often without requiring paired 3D supervision. Early
investigations of NeRF for X-ray angiography demonstrated the
feasibility of this strategy but also identified limitations arising
from sparse angular sampling, foreground--background ambiguity,
vessel overlap, and the sparse occupancy of vascular structures
\cite{maas2023nerf}. NeRF-CA subsequently introduced a static--dynamic
scene decomposition that models the moving coronary vasculature
separately from the background, enabling four-dimensional
reconstruction from angiographic sequences acquired at very few
viewing angles \cite{maas2025nerf}. NeCA represents coronary occupancy
using a multiresolution hash-encoded neural implicit field and
optimizes it through a differentiable cone-beam projector using two
projections \cite{wang2024neca}. 3DGR-CAR represents the coronary tree
using sparse 3D Gaussian primitives; a learned Gaussian-center
predictor supplies an initial point distribution, after which the
Gaussian parameters are refined through projection-based
optimization \cite{fu20243dgr}.

Patient-specific optimization promotes agreement with the acquired
measurements and reduces reliance on direct prediction of the complete
anatomy. Nevertheless, it requires iterative optimization at inference,
which can be computationally demanding. Moreover, projection
consistency alone does not guarantee a unique reconstruction from
ultra-sparse views, and the solution remains dependent on the imposed
priors and regularization. 3DGR-CAR mitigates this ambiguity through
learned initialization, but its Gaussian-center predictor requires
supervised training on paired projections and CT-derived 3D targets.

\subsection{Diffusion Priors for Medical Inverse Problems and Vascular Anatomy}
\label{sec:related_diffusion}

Score-based and diffusion generative models provide a principled mechanism for
incorporating learned anatomical information into inverse problems. A
generative model can first be trained to represent a measurement-independent
prior distribution $p(\boldsymbol{x})$ and subsequently combined with a known
forward model during inference \cite{song2021solving}. CCDF accelerates conditional diffusion
by perturbing an initial reconstruction to an intermediate
diffusion timestep and applying a shortened reverse process,
demonstrating how an informative initialization can reduce
the required sampling steps \cite{chung2022come}. Diffusion posterior
sampling introduces likelihood-derived guidance into the reverse process to
promote agreement with noisy or nonlinear measurements
\cite{chung2022diffusion}, while DiffusionMBIR extends this strategy to
3D medical reconstruction using pretrained diffusion priors
and model-based constraints \cite{chung2023solving}. Separating prior learning
from measurement inversion allows the same anatomical model to be reused
without training a new end-to-end reconstruction network for every acquisition
configuration.

Diffusion models have also been developed specifically to represent vascular
anatomy. VesselDiffusion generates three-dimensional vascular networks using a
two-stage formulation that combines two-dimensional vascular mask generation with
conditional 3D diffusion \cite{guo2025vesseldiffusion}. VesselFusion applies
conditional diffusion to vessel centerline extraction from an available 3D CT
volume and uses coarse-to-fine prediction with voting-based aggregation to
improve structural stability \cite{mita2026vesselfusion}. These studies
demonstrate that diffusion models can capture vascular continuity, branching,
and anatomical variability. Nevertheless, they address vascular generation or
extraction from existing three-dimensional data and do not enforce
patient-specific consistency with sparse angiographic projections. They
therefore establish the suitability of diffusion models for vascular
representation, but not how such a prior should be incorporated into a
severely underdetermined coronary reconstruction problem.

Within coronary reconstruction, XRA-Diff performs learned refinement of a
backprojected volume using a diffusion model conditioned on two segmented
angiographic projections and the corresponding backprojected representation
\cite{lau2026xradiff}. The anatomical prior and projection-conditioning
pathway are learned jointly from paired projection--volume examples, with
synthetic projections generated directly from 3D coronary masks rather than
CT intensity volumes. The basic configuration employs separate diffusion
models for the left and right coronary trees, although a model trained on
both sides is also evaluated. The model is trained with a two-view
conditioning representation, and reduced reconstruction accuracy is reported
when additional projections are used to construct the backprojected volume.
Patient-specific measurement information is incorporated through
backprojection and learned conditioning, without an explicit
projection-based data-consistency update during inference. Consequently,
the reported sampling procedure does not directly use reprojection error
to correct the reconstruction.

\begin{figure*}[t]
    \centering
    \includegraphics[width=\linewidth]{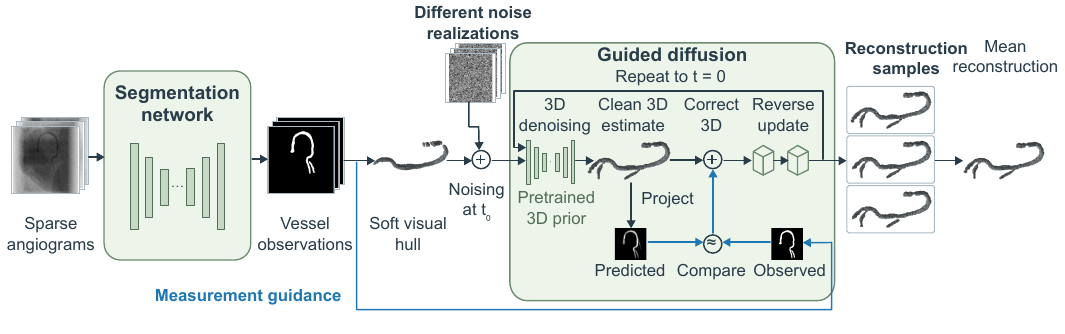}
\caption{Overview of the proposed sparse-view coronary reconstruction
framework. Angiograms are segmented to obtain vessel
observations for soft visual-hull initialization and projection
guidance. Starting from a noised hull, reconstruction uses a
pretrained 3D diffusion prior learned independently of projection
data. At each step, the estimated clean 3D volume is forward-projected
and compared with the observations to compute a correction before
the reverse-diffusion update. Different noise realizations produce
$L$ reconstruction samples, which are averaged voxelwise.}
    \label{fig:intro}
\end{figure*}

\section{Method}
\label{sec:method}

\subsection{Problem Formulation and Framework Overview}
\label{sec:problem_formulation}

We reconstruct the three-dimensional coronary vasculature relevant to
image-guided intervention, rather than the complete attenuation volume.
Given $N$ X-ray angiographic images
$\mathcal{I}=\{\boldsymbol{I}_i\}_{i=1}^{N}$ acquired at known projection
geometries $\mathcal{G}=\{\boldsymbol{P}_i\}_{i=1}^{N}$, the objective is to
estimate a vessel occupancy volume
$\boldsymbol{x}\in[0,1]^{H\times W\times D}$.

The proposed framework proceeds in three stages, as shown in Figure~\ref{fig:intro}. First, vessel structures are
extracted from the X-ray images to obtain a set of 2D vessel observations.
Second, these observations are geometrically backprojected to construct a
visual-hull estimate that provides rapid coarse localization of the vessel in
3D. Since sparse projections alone do not uniquely determine the underlying
vascular structure, the resulting inverse problem is subsequently addressed
within a Bayesian framework, where projection-based data consistency provides
the measurement constraint and a learned 3D diffusion model provides an
anatomical prior.

\subsection{Vessel Extraction from X-ray Angiography}
\label{sec:vessel_extraction}

For each X-ray angiographic image $\boldsymbol{I}_i$, $i=1,\ldots,N$,
a segmentation model $\boldsymbol{S}_{\phi}$ with parameters $\phi$
predicts a vessel probability map
$\boldsymbol{s}_i=\boldsymbol{S}_{\phi}(\boldsymbol{I}_i)
\in[0,1]^{H_d\times W_d}$, where $H_d$ and $W_d$ denote the height
and width of the projection images, respectively. These maps form
the collection of vessel observations
$\mathcal{S}=\{\boldsymbol{s}_i\}_{i=1}^{N}$.
The corresponding binary vessel mask is obtained by thresholding:
\begin{equation}
    \boldsymbol{m}_i
    =
    \mathbb{I}\left(\boldsymbol{s}_i\geq\tau_{\mathrm{seg}}\right)
    \in\{0,1\}^{H_d\times W_d},
\end{equation}
where $\tau_{\mathrm{seg}}$ is the segmentation threshold and
$\mathbb{I}(\cdot)$ denotes the elementwise indicator function.

\subsection{Fast Geometric Initialization with Visual Hull}
\label{sec:visual_hull}

To obtain an initial three-dimensional estimate directly from the segmented
projections, we employ visual-hull reconstruction. For each acquisition view,
the binary vessel support $\boldsymbol{m}_i$ is backprojected through the
corresponding imaging geometry. Denoting the backprojection operation
associated with geometry $\boldsymbol{P}_i$ by $\mathcal{B}_i$, the
three-dimensional support induced by the $i$-th observation is

\begin{equation}
    \mathcal{V}_i
    =
    \mathcal{B}_i(\boldsymbol{m}_i).
\end{equation}

Each $\mathcal{V}_i$ contains the voxels that are geometrically compatible with
the vessel silhouette observed in the corresponding projection. The visual
hull is then obtained by intersecting these supports across all available
views,

\begin{equation}
    \boldsymbol{x}_{\mathrm{VH}}
    =
    \bigcap_{i=1}^{N}
    \mathcal{V}_i .
\end{equation}

The visual hull provides a computationally efficient estimate of the spatial
region in which the vessel may be located. It is constructed directly from
the segmented projections and the calibrated acquisition geometry.

\subsection{Bayesian Formulation of the Sparse-view Inverse Problem}
\label{sec:bayesian_formulation}

Given the vessel observations $\mathcal{S}$ and known acquisition geometry
$\mathcal{G}$, Bayes' rule expresses the posterior distribution as
\begin{equation}
    p(\boldsymbol{x}\mid\mathcal{S},\mathcal{G})
    \propto
    p(\mathcal{S}\mid\boldsymbol{x},\mathcal{G})\,p(\boldsymbol{x}),
    \label{eq:bayesian_posterior}
\end{equation}
where the likelihood $p(\mathcal{S}\mid\boldsymbol{x},\mathcal{G})$
quantifies the consistency of a candidate vessel volume with the observed
projections, and the prior $p(\boldsymbol{x})$ encodes knowledge of
plausible three-dimensional coronary anatomy.

This decomposition provides the theoretical motivation for incorporating a
prior into sparse-view vessel reconstruction. As the number of observations
decreases, the likelihood alone admits an increasingly large set of
measurement-compatible solutions. The prior provides complementary structural
information that is not recoverable from the available projections alone.

\subsection{Projection Observation Model and Data Consistency}
\label{sec:data_consistency}

We next define the relationship between a candidate vessel occupancy
volume $\boldsymbol{x}$ and its two-dimensional vessel projections.
For acquisition geometry $\boldsymbol{P}_i$, let $\boldsymbol{A}_i$
denote the corresponding discrete cone-beam projection operator.
The predicted soft vessel silhouette is
\begin{equation}
    \widehat{\boldsymbol{s}}_i
    =
    \mathcal{A}_i(\boldsymbol{x})
    =
    \boldsymbol{1}
    -
    \exp\left(
        -\frac{
            \boldsymbol{A}_i
            [-\log(\boldsymbol{1}-\boldsymbol{x}_{\epsilon})]
        }{c}
    \right),
    \label{eq:vessel_forward_operator}
\end{equation}
where
$\boldsymbol{x}_{\epsilon}
=\operatorname{clip}(\boldsymbol{x},\epsilon,1-\epsilon)$
ensures numerical stability, and $c$ is the smallest effective
voxel spacing. The logarithm and exponential are applied
elementwise. This mapping converts the three-dimensional occupancy
field into a soft vessel silhouette with values in $[0,1]$.

The predicted silhouettes are compared with the binary vessel masks
$\boldsymbol{m}_i$ defined in Section~\ref{sec:vessel_extraction}.
The segmentation probability maps $\boldsymbol{s}_i$ additionally
provide pixelwise confidence weights, allowing uncertain
segmentation predictions to contribute less to the objective.
We define the data-consistency term using confidence-weighted
binary cross-entropy:
\begin{equation}
    \mathcal{L}_{\mathrm{DC}}(\boldsymbol{x})
    =
    \frac{1}{Z}
    \sum_{i=1}^{N}\sum_u
    \omega_{i,u}\,
    \operatorname{BCE}
    \left(
        \widehat{s}_{i,u},
        m_{i,u}
    \right),
    \label{eq:data_consistency}
\end{equation}
where $u$ indexes detector pixels, $\omega_{i,u}$ is the confidence
weight derived from $s_{i,u}$, and
$Z=\max(\sum_{i,u}\omega_{i,u},1)$ normalizes the total weight
across pixels and views. The predicted silhouette values are
clipped away from zero and one when evaluating the loss. This objective measures agreement between the candidate vessel
geometry and the segmented observations.

\subsection{3D Coronary Diffusion Prior}
\label{sec:diffusion_prior}

To constrain the sparse-view inverse problem toward anatomically plausible
solutions, we learn a prior distribution over three-dimensional coronary
vessel structures using a diffusion model.

Let
\begin{equation}
    \boldsymbol{x}_0
    \sim
    p_{\mathrm{data}}(\boldsymbol{x})
\end{equation}
denote a clean 3D vessel occupancy volume sampled from the training
distribution. Before diffusion, the volume is normalized as
$\boldsymbol{z}_0=2\boldsymbol{x}_0-\boldsymbol{1}$.
The forward diffusion process progressively perturbs
$\boldsymbol{z}_0$ according to
\begin{equation}
    q(
        \boldsymbol{z}_t
        \mid
        \boldsymbol{z}_0
    )
    =
    \mathcal{N}
    \left(
        \sqrt{\bar{\alpha}_t}\boldsymbol{z}_0,
        (1-\bar{\alpha}_t)\boldsymbol{I}
    \right),
    \label{eq:forward_diffusion}
\end{equation}
where $\bar{\alpha}_t=\prod_{k=1}^{t}(1-\beta_k)$ and
$\{\beta_k\}$ specifies the diffusion noise schedule.
Equivalently,
\begin{equation}
    \boldsymbol{z}_t
    =
    \sqrt{\bar{\alpha}_t}\boldsymbol{z}_0
    +
    \sqrt{1-\bar{\alpha}_t}\boldsymbol{\epsilon},
    \qquad
    \boldsymbol{\epsilon}
    \sim
    \mathcal{N}(\boldsymbol{0},\boldsymbol{I}).
    \label{eq:diffusion_reparameterization}
\end{equation}

A neural network
$\boldsymbol{\epsilon}_{\theta}(\boldsymbol{z}_t,t)$
is trained to predict the added noise by minimizing
\begin{equation}
    \mathcal{L}_{\mathrm{diff}}
    =
    \mathbb{E}_{
        \boldsymbol{x}_0,t,\boldsymbol{\epsilon}
    }
    \left[
        \left\|
            \boldsymbol{\epsilon}
            -
            \boldsymbol{\epsilon}_{\theta}
            (\boldsymbol{z}_t,t)
        \right\|_2^2
    \right].
    \label{eq:diffusion_loss}
\end{equation}

Through this training process, the diffusion model learns an approximation
to the distribution of three-dimensional vessel structures represented
in the training population.
The role of the diffusion model can be related to the Bayesian
formulation. At diffusion time $t$, the predicted noise provides an
estimate of the score of the perturbed prior distribution:
\begin{equation}
    \nabla_{\boldsymbol{z}_t}
    \log p_t(\boldsymbol{z}_t)
    \approx
    -
    \frac{
        \boldsymbol{\epsilon}_{\theta}(\boldsymbol{z}_t,t)
    }{
        \sqrt{1-\bar{\alpha}_t}
    },
    \label{eq:diffusion_prior_score}
\end{equation}
where $p_t$ denotes the distribution of the normalized vessel volumes
after perturbation at timestep $t$.

This motivates a guided reconstruction process in which the diffusion
model promotes anatomically plausible vessel structures, while the
data-consistency term encourages agreement with the observed vessel
projections. These components play complementary roles, consistent
with the Bayesian motivation in Equation~\eqref{eq:bayesian_posterior}.

\subsection{Visual-Hull-Initialized Diffusion Reconstruction}
\label{sec:posterior_reconstruction}

The preceding components are combined during reconstruction by using
the visual hull as a measurement-derived initialization for
diffusion-based refinement. The soft visual-hull initialization is
defined as
\begin{equation}
    \boldsymbol{c}_{\mathrm{VH}}
    =
    \lambda_{\mathrm{VH}}\boldsymbol{x}_{\mathrm{VH}},
    \label{eq:vh_soft_condition}
\end{equation}
where $\boldsymbol{x}_{\mathrm{VH}}$ denotes the binary visual-hull
volume and $0<\lambda_{\mathrm{VH}}\leq1$ scales the occupancy assigned
to its support.

A larger $\lambda_{\mathrm{VH}}$ assigns higher initial occupancy
values within the visual hull. The visual hull therefore provides
measurement-derived spatial localization without being treated as
an anatomical prior.

For a selected starting diffusion step $t_0$, the initialization
is converted to diffusion coordinates and perturbed according to
\begin{equation}
    \boldsymbol{z}_{t_0}
    =
    \sqrt{\bar{\alpha}_{t_0}}
    \left(
        2\boldsymbol{c}_{\mathrm{VH}}-\boldsymbol{1}
    \right)
    +
    \sqrt{1-\bar{\alpha}_{t_0}}\boldsymbol{\epsilon},
    \qquad
    \boldsymbol{\epsilon}
    \sim
    \mathcal{N}(\boldsymbol{0},\boldsymbol{I}).
    \label{eq:vh_diffusion_initialization}
\end{equation}

The parameter $t_0$ determines the initial noise level and the number
of reverse diffusion steps. A smaller $t_0$ preserves more of the
measurement-derived geometric structure in the initial state,
whereas a larger $t_0$ introduces stronger perturbations and a longer
reverse diffusion trajectory.

Starting from $\boldsymbol{z}_{t_0}$, reconstruction proceeds toward
$t=0$. At each step, the diffusion model first predicts a clean
vessel volume. Its corresponding occupancy volume is evaluated
using the projection-based data-consistency objective. Measurement
guidance is then applied to the predicted clean volume, which is
subsequently used to compute the next diffusion state.

Consequently, the proposed reconstruction combines three complementary
components: the visual hull provides coarse measurement-derived 3D
localization, the diffusion model provides a learned anatomical prior,
and the data-consistency term encourages agreement with the available
observations. This separation allows the visual hull to guide the
initial spatial localization while allowing the final vessel geometry
to deviate from the initial hull.

At diffusion step $t$, let
$\widehat{\boldsymbol{z}}_{0|t}$ denote the predicted clean volume
in normalized diffusion coordinates. Measurement guidance is
applied through
\begin{equation}
    \widehat{\boldsymbol{z}}_{0|t}^{+}
    =
    \operatorname{clip}
    \left(
        \widehat{\boldsymbol{z}}_{0|t}
        -
        \eta\widetilde{\boldsymbol{g}}_t,
        -1,1
    \right),
    \label{eq:denoised_dc_correction}
\end{equation}
where $\widetilde{\boldsymbol{g}}_t$ is the normalized,
backprojection-based correction computed from the
data-consistency objective Equation~\ref{eq:data_consistency} and expressed in diffusion coordinates.
The parameter $\eta>0$ controls the correction strength.
The clipping operation limits each voxel value to $[-1,1]$,
consistent with the normalization
$\boldsymbol{z}=2\boldsymbol{x}-\boldsymbol{1}$.
The adjusted estimate
$\widehat{\boldsymbol{z}}_{0|t}^{+}$ is subsequently used
in the Denoising Diffusion Probabilistic Models (DDPM) \cite{ho2020denoising} reverse update $\mathcal{R}$ to obtain
$\boldsymbol{z}_{t-1}$.
The complete procedure is summarized in
Algorithm~\ref{alg:vh_posterior}.

\begin{algorithm}[t]
\caption{Visual-Hull-Initialized Diffusion Reconstruction}
\label{alg:vh_posterior}
\begin{algorithmic}[1]

\Require
Vessel observations $\mathcal{S}=\{\boldsymbol{s}_i\}_{i=1}^{N}$
and corresponding binary masks $\{\boldsymbol{m}_i\}_{i=1}^{N}$;
projection geometries $\mathcal{G}=\{\boldsymbol{P}_i\}_{i=1}^{N}$;
soft visual-hull initialization $\boldsymbol{c}_{\mathrm{VH}}$;
pretrained diffusion model $\boldsymbol{\epsilon}_{\theta}$;
diffusion schedule $\{\bar{\alpha}_t\}$;
starting diffusion step $t_0$;
measurement-guidance strength $\eta$.

\Ensure
Reconstructed 3D vessel volume $\widehat{\boldsymbol{x}}$.

\State Sample
$\boldsymbol{\epsilon}\sim\mathcal{N}(\boldsymbol{0},\boldsymbol{I})$.

\State Initialize the diffusion state from the soft visual hull:
\[
    \boldsymbol{z}_{t_0}
    \gets
    \sqrt{\bar{\alpha}_{t_0}}
    (2\boldsymbol{c}_{\mathrm{VH}}-\boldsymbol{1})
    +
    \sqrt{1-\bar{\alpha}_{t_0}}\boldsymbol{\epsilon}.
\]

\For{$t=t_0,t_0-1,\ldots,1$}

    \State Predict the added noise:
    \[
        \widehat{\boldsymbol{\epsilon}}_t
        \gets
        \boldsymbol{\epsilon}_{\theta}(\boldsymbol{z}_t,t).
    \]

    \State Estimate the clean volume:
    \[
        \widehat{\boldsymbol{z}}_{0|t}
        \gets
        \operatorname{clip}
        \left(
            \frac{
                \boldsymbol{z}_t
                -
                \sqrt{1-\bar{\alpha}_t}\,
                \widehat{\boldsymbol{\epsilon}}_t
            }{
                \sqrt{\bar{\alpha}_t}
            },
            -1,1
        \right).
    \]

    \State Convert to vessel occupancy:
    \[
        \widehat{\boldsymbol{x}}_{0|t}
        \gets
        \frac{
            \widehat{\boldsymbol{z}}_{0|t}+\boldsymbol{1}
        }{2}.
    \]

    \State Compute the normalized measurement correction
    $\widetilde{\boldsymbol{g}}_t$ from
    $\mathcal{L}_{\mathrm{DC}}
    (\widehat{\boldsymbol{x}}_{0|t})$.

    \State Apply measurement guidance to the predicted clean volume:
    \[
        \widehat{\boldsymbol{z}}_{0|t}^{+}
        \gets
        \operatorname{clip}
        \left(
            \widehat{\boldsymbol{z}}_{0|t}
            -
            \eta\widetilde{\boldsymbol{g}}_t,
            -1,1
        \right).
    \]

    \State Perform one reverse-diffusion update:
    \[
        \boldsymbol{z}_{t-1}
        \gets
        \mathcal{R}
        \left(
            \boldsymbol{z}_t,
            \widehat{\boldsymbol{z}}_{0|t}^{+},
            t
        \right).
    \]

\EndFor

\State Convert the final state to vessel occupancy:
\[
    \widehat{\boldsymbol{x}}
    \gets
    \operatorname{clip}
    \left(
        \frac{\boldsymbol{z}_0+\boldsymbol{1}}{2},
        0,1
    \right).
\]

\State \Return $\widehat{\boldsymbol{x}}$.

\end{algorithmic}
\end{algorithm}

\section{Experiments}
\label{sec:experimental_setup}

\subsection{Datasets and Preprocessing}
\label{sec:data}

Experiments were conducted using the ImageCAS dataset
\cite{zeng2023imagecas}, which provides 1,000 coronary CT angiography volumes from 1,000 patients. For this study, the left and right
coronary trees were treated as separate reconstruction samples.

The CT volumes and binary vessel masks were resampled to
a common grid of $128\times128\times128$ voxels while preserving the selected
physical field of view. The diffusion-prior dataset comprised 1,981 side-specific
coronary-tree samples, partitioned into 1,387 training,
297 validation, and 297 test samples. From the test
partition, 110 samples were selected for the reconstruction
study.

\subsection{Implementation Details}
\label{sec:implementation}

\subsubsection{Diffusion Prior Training}

An unconditional three-dimensional DDPM
was trained as a spatial prior over binary coronary occupancy volumes.
Training masks were resized to $128\times128\times128$ voxels, mapped linearly from $[0,1]$ to $[-1,1]$,
and augmented using independent reflections along the three spatial axes.

The noise-prediction network was a single-channel 3D U-Net with a base feature
width of 32 and two spatial downsampling stages. Residual blocks used group
normalization and SiLU activations, and diffusion time was represented using a
128-dimensional sinusoidal embedding. The forward diffusion process comprised $T=200$ timesteps with
a linear variance schedule from $\beta_1=10^{-4}$ to
$\beta_T=2\times10^{-2}$. The model was optimized
using the standard noise-prediction mean-squared-error objective in Equation~\ref{eq:diffusion_loss}.

Training used AdamW with a learning rate of $10^{-4}$, weight decay
$10^{-5}$, and batch size one for at most 1,000 epochs.

\subsubsection{Projection Simulation}

Synthetic contrast-enhanced radiographs were generated using a differentiable
cone-beam projector implemented in Tomosipo \cite{hendriksen2021tomosipo}. CT intensities were clipped to $[-1000,3000]$ HU and converted into a
background attenuation volume. A vessel-specific attenuation increment
$\Delta\mu$ was then added within the regions defined by the corresponding
binary vessel masks to simulate iodine contrast enhancement. Quantum
noise was simulated by applying Poisson sampling in the transmission domain,
after which each projection was independently normalized to $[0,1]$. The
simulation modeled monochromatic Beer--Lambert attenuation and Poisson noise \cite{hansen2021computed}.

\subsubsection{Vessel Segmentation}

Projection data for segmentation training were generated using eight
uniformly distributed views over $360^\circ$. The source-to-isocenter and
isocenter-to-detector distances were 350 and 400~mm, respectively. A
$512\times512$ detector covered a physical area of
$300\times300~\mathrm{mm}^2$. The photon count was $10^5$, and the vessel
contrast was fixed at $\Delta\mu=0.008$. The resulting projection images were
resized to $256\times256$ pixels before segmentation training.

Vessel segmentation was performed using a three-level 2D U-Net with a base
feature width of 32. The network was trained for 100 epochs using the
unweighted sum of binary cross-entropy and soft Dice losses. AdamW was used
with a learning rate of $10^{-3}$, weight decay $10^{-5}$, and batch size
eight. Data augmentation comprised independent horizontal and vertical
reflections and additive Gaussian noise. The checkpoint achieving the highest
validation soft Dice was selected.

The trained network was applied to the reconstruction projections
without retraining. Its sigmoid output yielded a vessel probability
map $\boldsymbol{s}_i$ for each view $i$. The corresponding binary
masks $\boldsymbol{m}_i$, used for visual-hull construction and
data-consistency guidance, were obtained using
$\tau_{\mathrm{seg}}=0.5$.

\subsubsection{Reconstruction Settings}

A binary visual hull was constructed by intersecting the vessel
supports backprojected from all available views. A soft
initialization was then obtained by assigning an occupancy value
of 0.30 inside the hull and zero outside. This volume was mapped
to normalized diffusion coordinates and perturbed according to
the forward diffusion process at the selected starting timestep
$t_0$. The reconstruction procedure, which combines the learned
diffusion prior with projection-based measurement guidance,
is summarized in Algorithm~\ref{alg:vh_posterior}.

For each case and acquisition condition, the simulated
projections, segmentation outputs, and soft visual-hull
initialization were held fixed across all tested reconstruction
parameter settings.

The implementation and evaluation code are publicly available
in \url{https://github.com/tianyuan1wang/sparse_view_3d_vessel_reconstruction.git}.

\subsection{Evaluation Protocol}
\label{sec:evaluation}

Each reconstruction case corresponded to one side-specific
coronary tree. From the held-out diffusion test partition,
110 cases were selected for the reconstruction study.
Experiment~1 used 50 development cases to investigate and
select the reconstruction parameters of the proposed method.
Experiments~2 and~3 shared a separate evaluation cohort of
50 cases, each from a different source CT volume.
The remaining 10 cases were reserved and not used in the
reported experiments.

The holdout was defined at the coronary-tree level rather
than the source-CT level. Although the evaluated vessel masks
were excluded from diffusion-model training, other coronary
trees from the same source CT volumes could appear in
diffusion or segmentation training.

Thresholded reconstructions were evaluated using Dice and clDice. Dice quantified volumetric overlap, whereas
clDice assessed centerline and topological agreement. Average precision (AP) \cite{pedregosa2011scikit} was used to
evaluate the ability of sample disagreement to identify
reconstruction errors. Disagreement values were used as prediction
scores, with incorrectly classified voxels in the thresholded
reconstruction treated as the positive class. AP summarized
precision across recall levels without requiring a fixed
threshold on the disagreement scores.

To assess reconstruction robustness to variations in segmentation quality,
we applied the same segmentation model, trained at $\Delta\mu=0.008$,
across all contrast conditions without retraining. This setup represents
a simplified deployment scenario in which vessel contrast may differ
from that encountered during training. Applying the fixed model to
lower-contrast images may reduce segmentation accuracy, resulting in
less reliable vessel observations for reconstruction.

\subsection{Effect of Measurement Quality on Diffusion Refinement}
\label{sec:exp_contrast_diffusion}

We first examined whether the diffusion refinement depth required for
reconstruction depends on the quality of the measurement-derived
initialization. Vessel contrast was used as a controlled proxy for image quality, with
$\Delta\mu\in\{0.004,0.008\}$. For each contrast level, we evaluated reconstructions from $N_{\mathrm{view}}\in\{2,4,8\}$ uniformly spaced projection views over $[0^\circ,180^\circ)$, while varying the diffusion start timestep over $t_0\in\{10,15,20,25,30\}$. Based on a comparison of two heuristic threshold candidates on parameter-selection cohort, $\tau=0.05$ and $0.15$ were selected for
$\Delta\mu=0.004$ and $0.008$, respectively. These thresholds were
held fixed across diffusion depths and view counts.

As described in Section~\ref{sec:posterior_reconstruction}, increasing $t_0$
introduces a stronger perturbation of the soft visual-hull initialization
before reverse diffusion. A larger $t_0$ consequently allows the learned prior
greater freedom to modify the measurement-derived geometry, whereas a smaller
$t_0$ retains more information from the initial visual hull. We therefore
interpret $t_0$ as the degree of diffusion-based refinement away from the
measurement-derived initialization, rather than as an explicit scalar weight
on the prior.

\begin{figure}[t]
    \centering
    \includegraphics[width=\linewidth]{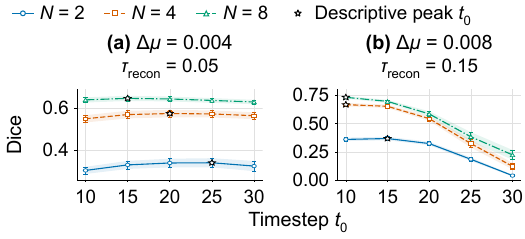}
    \caption{Effect of the diffusion start timestep $t_0$ on
    fixed-threshold 3-D Dice (a,b) for
    $\Delta\mu=0.004$ and $\Delta\mu=0.008$. Curves correspond to
    $N_{\mathrm{view}}\in\{2,4,8\}$, using reconstruction thresholds of
    $\tau_{\mathrm{recon}}=0.05$ and $0.15$ for the two contrast levels,
    respectively.  Stars indicate the value of $t_0$ attaining
    the highest observed mean for the corresponding metric and view count
    within the evaluated grid.}
    \label{fig:exp1_main}
\end{figure}

Figure~\ref{fig:exp1_main} shows that the influence of $t_0$ depended on both
measurement quality and the evaluation criterion. At the lower contrast
$\Delta\mu=0.004$, fixed-threshold Dice exhibited broad, shallow maxima
[Figure~\ref{fig:exp1_main}(a)]. The highest observed mean shifted from
$t_0=25$ for two views to $t_0=20$ for four views and $t_0=15$ for eight
views. This trend is consistent with stronger prior-mediated correction being
useful when angular information is limited: fewer views produce a less
constrained visual hull and therefore permit greater benefit from departing
from the initial geometry. Nevertheless, the differences between adjacent
values of $t_0$ were modest, particularly for four and eight views. The
separation between the view-count curves was substantially larger than the
variation within each curve, indicating that angular coverage was the
dominant factor under low-contrast conditions.

A different behavior was observed at the higher contrast
$\Delta\mu=0.008$. It deteriorated markedly as $t_0$
increased beyond the shallowest settings
[Figure~\ref{fig:exp1_main}(b)]. The highest observed Dice occurred at
$t_0=15$ for two views and at $t_0=10$ for four and eight views. The decrease was
particularly pronounced for $t_0\geq20$, showing that excessive perturbation
of an already informative measurement-derived initialization can remove
vessel structures supported by the observations and degrade both volumetric
overlap and centerline consistency.

Overall, these results show that there is no universally optimal diffusion
start timestep. When the observations provide reliable geometric information,
the reconstruction should remain close to the visual-hull initialization and
only limited diffusion refinement is beneficial. Under lower-quality and
more angularly sparse measurements, greater refinement can improve volumetric
overlap. The preferred setting is therefore determined jointly by image
quality, angular coverage, and the reconstruction objective. For subsequent experiments, $t_0$ was fixed at $20$ for
$\Delta\mu=0.004$ and $10$ for $\Delta\mu=0.008$,
independent of the number of projection views.
The larger $t_0$ at lower contrast allows greater diffusion
refinement to compensate for less reliable vessel observations,
whereas the smaller $t_0$ at higher contrast better preserves
the more informative visual-hull initialization.

\subsection{Comparison with Measurement-Driven Reconstruction Methods}
\label{sec:exp_method_comparison}

We next evaluated whether incorporating a pretrained three-dimensional
coronary diffusion prior improves reconstruction relative to
measurement-only geometry and patient-specific neural regularization.
For the proposed method, we used the voxelwise mean of ten reconstructions
generated with different noise realizations. The proposed method was
compared with the visual hull (VH), deep image prior (DIP)
\cite{ulyanov2018deep}, and hash-grid implicit neural representation
(INR) \cite{wang2024neca,mescheder2019occupancy} on the matched
method-comparison cohort. 

The four methods use different types of prior information. VH directly
intersects the backprojected vessel silhouettes and therefore relies
only on measurement-derived geometry. DIP optimizes a randomly
initialized convolutional generator for each case, introducing an
architectural prior without population-level anatomical training.
Hash-grid INR encodes normalized 3D coordinates using a multiresolution
hash grid and predicts vessel occupancy through a neural network
optimized for each case. The proposed method combines patient-specific
projection consistency with a diffusion prior learned from a population
of three-dimensional coronary trees.

Reconstruction thresholds
were held fixed across view counts. For each method and contrast level, the reconstruction threshold was selected using the parameter-selection cohort and subsequently held fixed across all view counts. For the proposed method, DIP, and hash-grid INR, respectively, the selected thresholds were $(0.05,\,0.145,\,0.76)$ at $\Delta\mu=0.004$ and $(0.15,\,0.82,\,0.83)$ at $\Delta\mu=0.008$. The binary visual hull was
evaluated directly after the common connected-component filtering.
Dice and clDice were used to assess volumetric overlap and centerline
agreement, respectively.

\begin{figure}[t]
    \centering
    \includegraphics[width=\linewidth]
    {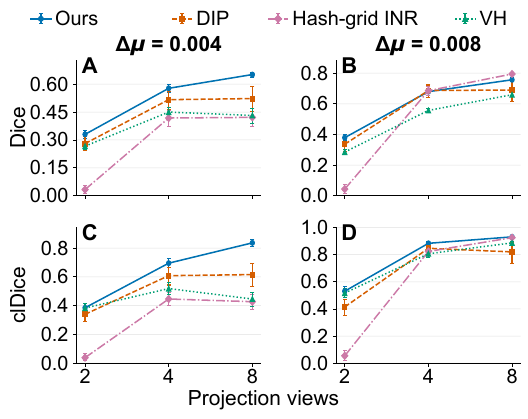}
    \caption{Comparison of the proposed method, DIP, hash-grid INR, and
    VH using fixed-threshold Dice (A,B) and clDice (C,D) at vessel
    contrasts $\Delta\mu=0.004$ and $\Delta\mu=0.008$. The proposed
    reconstruction is the voxelwise mean of ten samples. All methods
    used the same cases, projection views, and segmentation-derived
    observations. Method-specific thresholds were selected on
    development data and held fixed across
    $N_{\mathrm{view}}\in\{2,4,8\}$. Markers show mean scores across 50 held-out coronary-tree cases;
    error bars indicate 95\% confidence intervals.}
    \label{fig:exp2_main}
\end{figure}

As shown in Figure~\ref{fig:exp2_main}, reconstruction accuracy
generally improved with additional projection views, although
neither metric increased monotonically for every method. At the lower contrast,
$\Delta\mu=0.004$, the proposed method achieved the highest mean Dice
and clDice at all three view counts
[Figure~\ref{fig:exp2_main} (A,C)], indicating better volumetric overlap
and centerline agreement. DIP achieved higher Dice than VH at all view
counts, but higher clDice only with four and eight views. Hash-grid INR
had particularly low scores with two views and improved substantially
with additional views.

At the higher contrast, $\Delta\mu=0.008$, the proposed method
achieved the highest mean Dice with two views. The three neural
reconstruction methods achieved similar mean Dice with four views,
whereas hash-grid INR attained the highest mean Dice with eight
views [Figure~\ref{fig:exp2_main}(B)]. For clDice, the proposed
method achieved the highest mean with two and four views and a
similar mean to hash-grid INR with eight views
[Figure~\ref{fig:exp2_main}(D)]. Hash-grid INR remained limited
under two-view acquisition.

In a matched comparison on the same NVIDIA A40 GPU at $\Delta\mu=0.004$,
the proposed method ($L=10$) required a mean wall-clock time of
4.49~min per phantom across two, four, and eight views. This corresponds
to speedups of $3.87\times$ over DIP (17.37~min) and $15.86\times$
over hash-grid INR (71.26~min).

\begin{figure}[t]
    \centering
    \includegraphics[width=\linewidth]
    {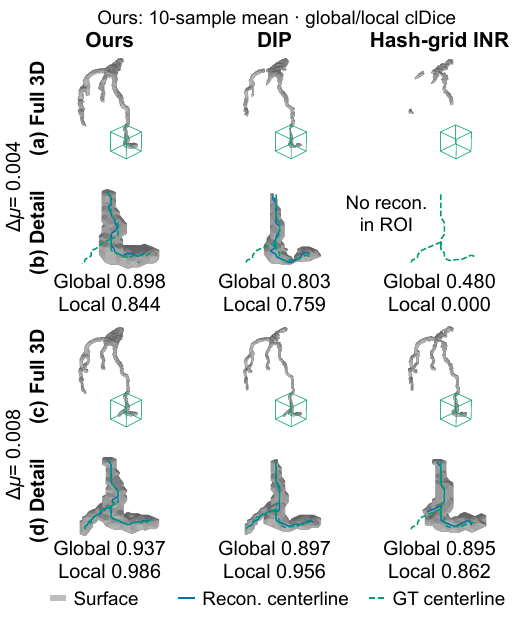}
    \caption{Reconstruction examples at $\Delta\mu=0.004$ (a,b) and
    $\Delta\mu=0.008$ (c,d). Rows (a,c) show the full 3D vessel
    structures, and rows (b,d) show enlarged regions marked by green
    boxes. Gray surfaces represent vessel geometry, solid blue lines
    show reconstruction-derived centerlines, and dashed green lines
    show ground-truth centerlines. Global clDice is reported for the
    complete reconstruction and local clDice for the highlighted
    region.}
    \label{fig:exp2_examples}
\end{figure}

Figure~\ref{fig:exp2_examples} shows reconstruction examples with eight views at both
contrast levels, together with enlarged regions comparing the
reconstructed and ground-truth centerlines. At low contrast, the
proposed method retained more of the branches in the highlighted region
than DIP and hash-grid INR, consistent with its higher global and local
clDice. At higher contrast, all three methods recovered these branches more completely. The proposed method achieved the highest global and local clDice scores, although the differences among methods were smaller than at lower contrast. These examples suggest that the advantage of the proposed method is more pronounced under reduced contrast, particularly in preserving vessel branches.

To assess the contribution of projection guidance, we disabled
the consistency updates ($\eta=0$) while retaining visual-hull
initialization and matching all other settings.
As shown in Table~\ref{tab:guidance_ablation}, guidance improved
mean Dice and clDice across all conditions, with larger gains
at lower contrast and with more projection views. At low contrast, unguided reconstruction deteriorated with increasing
view count. This may reflect the sensitivity of visual-hull
intersection to false-negative segmentations: additional imperfect
views can remove valid vessel structures from the initialization,
and unguided diffusion lacks subsequent measurement-based correction.

\begin{table*}[t]
\caption{Projection-guidance ablation on the 50 held-out coronary-tree
cases from Section \ref{sec:exp_method_comparison}. Values are mean $\pm$ SD across cases,
using ten-sample mean reconstructions. All other settings are matched.}
\label{tab:guidance_ablation}
\centering
\small
\setlength{\tabcolsep}{5pt}   
\renewcommand{\arraystretch}{0.95}

\begin{tabular}{@{}cc*{4}{c}@{}}
\toprule
$\Delta\mu$ & Views
& \multicolumn{2}{c}{Dice $\uparrow$}
& \multicolumn{2}{c}{clDice $\uparrow$} \\
\cmidrule(lr){3-4}\cmidrule(lr){5-6}
& & No guidance & Full method & No guidance & Full method \\
\midrule
0.004 & 2 & $0.289 \pm 0.089$ & $0.330 \pm 0.085$
          & $0.338 \pm 0.119$ & $0.382 \pm 0.119$ \\
0.004 & 4 & $0.278 \pm 0.127$ & $0.578 \pm 0.075$
          & $0.331 \pm 0.153$ & $0.696 \pm 0.112$ \\
0.004 & 8 & $0.186 \pm 0.126$ & $0.652 \pm 0.048$
          & $0.219 \pm 0.145$ & $0.837 \pm 0.083$ \\
\addlinespace
0.008 & 2 & $0.367 \pm 0.069$ & $0.379 \pm 0.069$
          & $0.517 \pm 0.116$ & $0.534 \pm 0.118$ \\
0.008 & 4 & $0.612 \pm 0.055$ & $0.680 \pm 0.055$
          & $0.786 \pm 0.079$ & $0.884 \pm 0.061$ \\
0.008 & 8 & $0.624 \pm 0.053$ & $0.757 \pm 0.047$
          & $0.769 \pm 0.083$ & $0.933 \pm 0.034$ \\
\bottomrule
\end{tabular}
\end{table*}

\subsection{Reconstruction Sampling for Accuracy and Error Localization}

\begin{figure}[t]
    \centering
    \includegraphics[width=\linewidth]{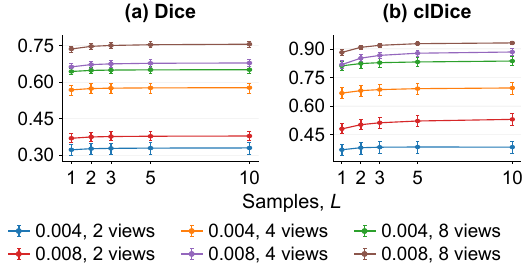}
    \caption{Effect of the number of reconstruction samples $L$
    on the Dice (a) and clDice (b) of the voxelwise mean reconstruction
    for two, four, and eight projection views at vessel contrasts
    $\Delta\mu\in\{0.004,0.008\}$.}
    \label{fig:exp3_mc}
\end{figure}

The proposed method generates multiple reconstructions using
different diffusion noise realizations. Whereas the preceding
comparison used the voxelwise mean of $L=10$ samples,
this experiment examines the effect of sample averaging
on reconstruction accuracy and the ability of sample
disagreement to identify reconstruction errors.
As shown in Figure~\ref{fig:exp3_mc}, mean Dice and clDice
generally increased with the number of samples, with
diminishing improvements beyond $L=5$.

\begin{table*}[!t]
\centering
\caption{Error localization using disagreement among $L=10$ reconstruction
samples. Values are means across source CT volumes
[95\% confidence intervals].}
\label{tab:exp3_error_localization}

\begin{threeparttable}
\setlength{\tabcolsep}{5pt}   
\renewcommand{\arraystretch}{0.95}

\begin{tabular}{@{}ccccc@{}}
\toprule
Contrast $\Delta\mu$
& Views
& Error prevalence
& Error-localization AP
& AP gain \\
\midrule
$0.004$ & 2 & $0.2353\,[0.2271,\,0.2437]$ & $0.4470\,[0.4281,\,0.4639]$ & $+0.2117\,[0.1973,\,0.2247]$ \\
$0.004$ & 4 & $0.1673\,[0.1618,\,0.1728]$ & $0.3226\,[0.3089,\,0.3362]$ & $+0.1553\,[0.1448,\,0.1657]$ \\
$0.004$ & 8 & $0.1510\,[0.1471,\,0.1548]$ & $0.2929\,[0.2829,\,0.3034]$ & $+0.1419\,[0.1335,\,0.1507]$ \\
$0.008$ & 2 & $0.2604\,[0.2500,\,0.2705]$ & $0.4060\,[0.4005,\,0.4115]$ & $+0.1456\,[0.1399,\,0.1517]$ \\
$0.008$ & 4 & $0.1494\,[0.1410,\,0.1580]$ & $0.3505\,[0.3459,\,0.3551]$ & $+0.2011\,[0.1933,\,0.2091]$ \\
$0.008$ & 8 & $0.1126\,[0.1057,\,0.1196]$ & $0.3261\,[0.3200,\,0.3318]$ & $+0.2135\,[0.2067,\,0.2205]$ \\
\bottomrule
\end{tabular}

\end{threeparttable}
\end{table*}

To assess error localization, we computed voxelwise disagreement among
$L=10$ reconstruction samples without using ground-truth information.
The evaluation region was defined by dilating the union of the
ground-truth mask and the thresholded $L=10$ sample-mean reconstruction
by two voxels using six-connectivity, without component filtering.

Within this region, incorrectly classified voxels were treated as the
positive class, and disagreement values were used as prediction scores.
Error-localization performance was quantified using AP. Error prevalence
was defined as the fraction of incorrectly classified voxels within the
evaluation region, and AP gain was calculated as AP minus error
prevalence, corresponding to the improvement over the constant-score
baseline. Metrics were first averaged across coronary-tree cases derived
from the same source CT volume. Cohort means were then calculated
across the 50 source CT volumes, and 95\% confidence intervals
were estimated by percentile bootstrapping over these volumes.

As reported in Table~\ref{tab:exp3_error_localization},
mean AP gains ranged from $0.1419$ to $0.2135$ across
the six contrast--view conditions. All corresponding
95\% confidence intervals were above zero, indicating
that sample disagreement provided information about
reconstruction-error locations within the evaluated regions.
These results support its potential as a spatial indicator
of unreliable reconstruction regions.

\section{Discussion}

This work combines a measurement-independent 3D coronary prior
with segmentation-derived observations through visual-hull
initialization and projection guidance. Unlike projection-conditioned
models learned from paired examples \cite{wang2025deepca},
anatomical-prior training requires only 3D vessel masks.
Patient-specific observations enter during reconstruction,
following the separation of prior learning and inversion
\cite{song2021solving,chung2022diffusion}. The same prior was
used across the evaluated view counts without retraining.

The low-contrast advantage over DIP and hash-grid INR is
consistent with anatomical regularization helping constrain
geometry when observations are ambiguous. At higher contrast,
the advantage was less consistent, and hash-grid INR achieved
higher Dice with eight views. More informative observations
may reduce the relative benefit of population-level regularization.
The limited two-view performance nevertheless shows that
accurate recovery remains challenging under severe angular sparsity.

The truncated reverse trajectory applies anatomical information
encoded in the trained denoiser. Increasing $t_0$ introduces both
greater perturbation and additional refinement steps; it does not
independently control a scalar prior weight. The deterioration
at higher contrast suggests that excessive perturbation can
compromise the benefit of an informative initialization.

At low contrast, unguided deterioration with additional views
may reflect the sensitivity of hull intersection to segmentation
false negatives. Additional views provide more constraints but
also more opportunities to exclude valid vessel support.
Projection updates permit the reconstruction to depart from
this initialization. The matched ablation isolates their benefit,
although the specific contribution of hull errors to the
unguided decline was not directly measured.

Sample averaging improved accuracy with diminishing returns,
consistent with reduced sampling variability. Disagreement also
identified error-prone regions retrospectively and could help
prioritize additional verification. However, averaging cannot
remove shared errors, and samples may agree on an incorrect
structure. Disagreement is therefore an error-localization cue
within the evaluated region, not a calibrated error probability;
low disagreement does not guarantee correctness.

The matched runtime advantage is relevant to preparing a 3D
reference before guidewire navigation
\cite{ravigopal2022fluoroscopic}. Rapid updates would require
further acceleration. The practical value of both the geometry
and disagreement maps remains to be established in navigation tasks.

Evaluation was limited to static, simulated ImageCAS anatomy
\cite{zeng2023imagecas}, uniformly spaced views, and a
$128^3$ reconstruction grid. Clinical angiograms, motion, and
fine-vessel accuracy in physical units require further assessment.
Mismatch between the surrogate observation model and segmented
projections may also affect guidance. Although the prior remained
fixed, $t_0$ was selected by acquisition condition and thresholds
were selected per contrast, then held fixed across view counts.
Reliable parameter selection under variable clinical observation
quality therefore remains an unresolved deployment requirement.

\section{Conclusion}
We presented a sparse-view 3D coronary reconstruction framework
that combines a diffusion prior learned from 3D vessel masks with
soft visual-hull initialization and projection-based data
consistency. The framework reconstructs coronary geometry from
segmentation-derived vessel observations. Experiments on simulated
angiographic data demonstrated improved reconstruction accuracy
at low contrast relative to the evaluated baselines. Averaging
reconstruction samples improved accuracy, while disagreement
among samples provided a spatial indicator of potentially
unreliable regions. Further validation on clinical angiograms
and in navigation tasks is needed to assess the framework's
potential to provide anatomical guidance for PCI.

\section*{REFERENCES}

\bibliographystyle{IEEEtran}
\bibliography{reference}

\end{document}